\documentclass[11pt, fullpage]{elsarticle}
\newcommand{\mypackages}{%
  \usepackage{amssymb}
  \usepackage{amsmath}
  \usepackage{amsthm}
  \usepackage[colorlinks=true,allcolors=red]{hyperref}
  \usepackage{graphicx}
  \usepackage{enumitem}
  \usepackage{changes}
  \usepackage{subfigure}
  \usepackage{soul}%added to enable highlighting \hl
\usepackage{xfrac}%defines sfrac
\usepackage{mathrsfs}
\graphicspath{{Figures/}{figures/}}
}
\mypackages

\usepackage{amssymb}

\graphicspath{{Figures/}{figures/}}
\newcounter{bla}

\journal{Computer Physics Communications}

\begin{document}

\begin{frontmatter}



\title{tda-segmentor: A tool to extract and analyze local structure and porosity features in porous materials}


\author[label1]{Aditya Vasudevan\corref{cor1}}
\author[label1]{Jorge Zorrilla Prieto}
\author[label1,label2]{Sergei Zorkaltsev}
\author[label1]{Maciej Haranczyk\corref{cor2}}

\cortext[cor1]{Corresponding author's email: adityavv.iitkgp@gmail.com}
\cortext[cor2]{Corresponding author's email: maciej.haranczyk@imdea.org}

\affiliation[label1]{organization={IMDEA Materials Institute},%Department and Organization
            addressline={Eric Kandel 2, Tecnogetafe}, 
            city={Madrid},
            postcode={28906}, 
            %state={Madrid},
            country={Spain}}

\affiliation[label2]{organization={Universidad Carlos III de Madrid},%Department and Organization
            addressline={Av. Universidad 30}, 
            city={Leganes, Madrid},
            postcode={28911}, 
            %state={Madrid},
            country={Spain}}
            
\begin{abstract}
Local geometrical features of a porous material such as the shape and size of a pore or the curvature of a solid ligament 
do often affect the macroscopic properties of the material, and their characterization is necessary to fully understand the structure-property relationships.
In this contribution, we present an approach to automatically segment large porous structures into such local features. Our work takes inspiration from techniques available in Topological Data Analysis (TDA). In particular, using Morse theory, we generate Morse-Smale
Complexes of our structures that segment the structure, and/or its porosity into individual features that can then be compared. We develop a tool 
written in C\texttt{++} that is built on the topology toolkit (TTK) library, an open source platform for the topological analysis of scalar data, with which we can perform segmentation of these structures. Our tool takes a volumetric grid representation as an input, which can be generated from atomistic or mesh structure models and any function defined on such grid, e.g. the distance to the surface or the interaction energy with a probe. We demonstrate the applicability of the tool by two examples related with analysis of porosity in zeolite materials as well as analysis of ligaments in a porous metal structure. Specifically, by segmenting the pores in the structure we demonstrate some applications to zeolites such as assessing pore-similarity between structures 
or evaluating the accessible volume to a target molecule such as methane that can be adsorbed to its surface. Moreover, once the Morse-Smale complexes are generated, we can construct graph representations of the void space, replacing the entire pore structure by a simply connected graph. Similarly, the same tool is used to segment and generate graphs representing the solid structure and we show how they can be used to correlate 
structure and mechanical properties of the material. The code is published as open-source and can be accessed here : \href{https://github.com/AMDatIMDEA/tda-segmentor}{https://github.com/AMDatIMDEA/tda-segmentor}
\end{abstract}


\begin{keyword}

structure-property relationships \sep pore segmentation \sep Topological Data Analysis (TDA) \sep Morse-Smale Complexes \sep Reeb Graphs



\end{keyword}

\end{frontmatter}

% \linenumbers

% All CPiP articles must contain the following
% PROGRAM SUMMARY.

{\bf PROGRAM SUMMARY}
%Delete as appropriate.

\begin{small}
	\noindent \\
	{\em Program Title: \texttt{tda-segmentor}}                                          \\ \\
	{\em CPC Library link to program files:} (to be added by Technical Editor) \\ \\
	{\em Developer's repository link:} \href{https://github.com/AMDatIMDEA/tda-segmentor}{https://github.com/AMDatIMDEA/tda-segmentor} \\ \\
	{\em Code Ocean capsule:} (to be added by Technical Editor)\\ \\
	{\em Licensing provisions:} BSD 3-clause \\ \\
	{\em Programming language:}  \texttt{C++}                                 \\ \\
	{\em Nature of problem:} Porous material properties such as gas adsorption in zeolites or MOF's, to macroscopic mechanical properties of gold nanostructures derive largely from the local geometric features of the pore or structure and understanding this relationship is quite important. Computational strategies have been extremely successful in this regard, and coupled with machine learning techniques can predict material properties trained on a small set of simulated data. However this requires robust descriptors to be generated that can be quite challenging. \\ \\
	{\em Solution method:} 	We present here a method and tool based on topological data analysis (TDA) that can analyze data purely based on its shape. From a volumetric grid representation of the structure, such as the distance grid or the energy grid to a probe molecule, we generate Morse-Smale complexes of the structures that segment the structure/pore into distinct segments. These can then be used to assess pore/structure similarity, develop robust descriptors, correlate and predict better the pore/structure-property relationship of the porous material. \\ 

\end{small}

\section{Introduction}
\label{sec:introduction}

Porous materials of various chemical compositions and pore size-scales are both commonly observed in nature and applied in technology, and have long attracted researchers' attention. For example, nanoporous materials such as zeolites, natural clays and metal organic frameworks (MOF) are finding applications in the industry as chemical catalysts \cite{KONNERTH2020213319}, membranes or as adsorbents \cite{Auerbach2003}. Porous metal structures are investigated in the context of lightweight structural materials for energy-efficient transport \cite{PATEL201820391}. Numerous others are investigated in the context of battery electrodes \cite{EGOROV2020201}, bio-materials \cite{zadpoor2019additively}, geological deposits and others.

Computational approaches have been playing a crucial role in understanding the structure-property relationships and aiding the design. For example, molecular simulations can be used to reliably predict adsorption and diffusion of guest species \cite{Smit2008, Krishna2007} . Similarly, in porous metals such as gold, simulations can shed light onto relationships between the topology of the structure and/or its pore network and the mechanical properties of the material. 
For instance, finite element simulations have shown that for two porous structures with same solid fractions but different pore topologies (gyroids vs spinodal-like)
 yield at different compressive stresses \citep{Mangipudi2016}. In addition through molecular dynamics simulations, compression 
 of nanoporous structures with the same solid fraction and ligament diameter but different topology are found to have different mechanical responses \citep{Mathesan2021, Mathesan2020}. 

Furthermore, in recent years a wide accessibility of high-throughput computing has allowed studies of larger sets of structures. Machine learning-based approaches have become to emerge, i.e. by training on a small set of simulated data, ML regressors can predict the material property just from the structure features (descriptors). For guest adsorption for example, porosity descriptors such as accessible surface area (ASA), largest 
cavity diameter (LCD) can readily generate feature vectors which can act as an input for the machine learning algorithms \cite{Simon2015}.  
However such pore descriptors only work well at high pressures where the guest molecule adsorbs in the entire void space, but for low pressures 
they tend to be localized in the strongly binding regions of the material’s pore \citep{Lin2012}. Hence there is a need to generate robust 
descriptors that is also able to include extremely local pore morphology of the structures too.

Topological Data Analysis (TDA) has already developed into a mature field that is capable of obtaining insight from large datasets 
by looking at their `shape' and how it can reveal important information of the data. In particular, persistent homology \citep{edelsbrunner2022} can explain how certain features of the data persist across multiple-scales. Such methods are already popular in network science \citep{Aktas2019} and have also
been recently used for applications in material science. In \citep{Lee2017}, persistent homology is used to formulate robust porosity descriptors 
and they show that two structures with similar pore morphology have so-called similar persistence diagrams. In \citep{Krishnapriyan2020}, such
persistence diagrams have been converted to persistence images which can then act as feature vectors for machine learning algorithms. 

Another topic of TDA is Morse Theory, where the topology of a manifold is analyzed by studying differentiable functions on that manifold. Given a scalar field, $f$, Morse theory, which in algebraic topology is used to analyze the topology of a manifold by studying differentiable functions on that manifold, cell-complexes of the manifold can be formed, which can be used in decomposing
the manifold into different unique regions. These segments are called Morse-Smale complexes and the idea is to gather some information of the function itself from the manifold segments. Such segmentation methods are extremely useful in applications where local features contribute to the property at much larger scales. For example, the segmentation of electron charge densities for a molecule can be used to calculate the atomic charge and other related properties \citep{Bhatia2018}; 3D X-ray computed tomography images (CT) \citep{Pandey2021} can be segmented to identify the grain structure; and invariants in the distance field can be identified for the analysis of porous structures \citep{Gyulassy2007}.  

Our approach in this work is based on the segmentation of a scalar field such as the distance field or the energy field for the structure and extract topological information from this field. However, we also highlight, other methods exist in TDA to analyze such structures, such as alpha complexes that can directly analyze a point cloud of atoms/structure without defining a scalar field over it and have achieved great results in the analysis of proteins \citep{Edelsbrunner1994, Masood2015, Liang1998-bv}. 
In this work, we develop a high-throughput tool written in C\texttt{++} that is built on the topology toolkit library (TTK) \citep{ttk}, which
has efficient algorithms for the topological analysis of scalar functions, and we build on these algorithms to analyze our data. The rest of the article is organized as follows: In section \ref{sec:methods}, we first present the TDA method to generate the segments. We discuss the different input fields that can be used for analysis, the methodology based on persistence used to filter noise and finally generate unique segments. In section \ref{sec:tda-segmentor} we apply it to some example zeolites and nanoporous gold structures to generate segments of either the pore or the structure. Further, using the segmentation information, we also demonstrate that we can replace the entire structure by a much simpler graph representation, and finally in section \ref{sec:applications}, we discuss some applications from the generated segments related to guest adsorption and pore-similarity. We conclude with some remarks in section \ref{sec:conclusion}.

%%%%%%%%%%%%%%%%%%%%%%%%%%%%%%%%%%%%%%%%%%%%%%%%%%%%%%%%%%%%%%%%%
\section{Methods}
\label{sec:methods}

Topological Data Analysis (TDA) is premised on the idea that the shape of data sets contains relevant information. Herein, we first
present some of the mathematical preliminaries of topological analysis that is based on persistent homology and we refer to \citep{Milnor, Forman2002, edelsbrunner2022} for an in-depth analysis of the theory. TDA has seen successful applications in a variety of fields of science such as combustion \citep{Bremer, Gyulassy2014}, astrophysics \citep{Shivashankar} and materials science \citep{Gyulassy2016, Lee2017}. The following subsections describe our implementation and the input data.

\subsection{Mathematical overview}
Let us consider a manifold $ \Gamma$ and a piece wise linear scalar field $ f : \Gamma \to \mathbb{R}$, where $ d $ is the dimension of the manifold $ \Gamma$.  While the theoretical framework can be extended to any dimension $ d$, we restrict ourselves here to $ d = 3$, which is the three dimensional Cartesian space. The function $ f $ has values at the vertices $ \Gamma^0$ of $ \Gamma$ and for higher dimension simplices, such as edges, faces or the volume within the faces, the function $ f$ is linearly interpolated. The first step is to evaluate the critical points of the scalar function $ f$. For smooth functions, the critical points are simply the points where the gradient of the 
scalar field, $ \nabla f $ vanishes, while in the discrete setting, the critical points are evaluated slightly differently (see \citep{ttk}). Once the critical points are evaluated, 
they are given indices ($\mathcal{I}$) with $ \mathcal{I} = 0$ for minima, $ \mathcal{I} = 1$ for 1-saddles, $ \mathcal{I} = 2$ for 2-saddles and $ \mathcal{I} = 3$ for maxima.

\begin{figure}[ht!]%
\centering%
\includegraphics[width = \textwidth]{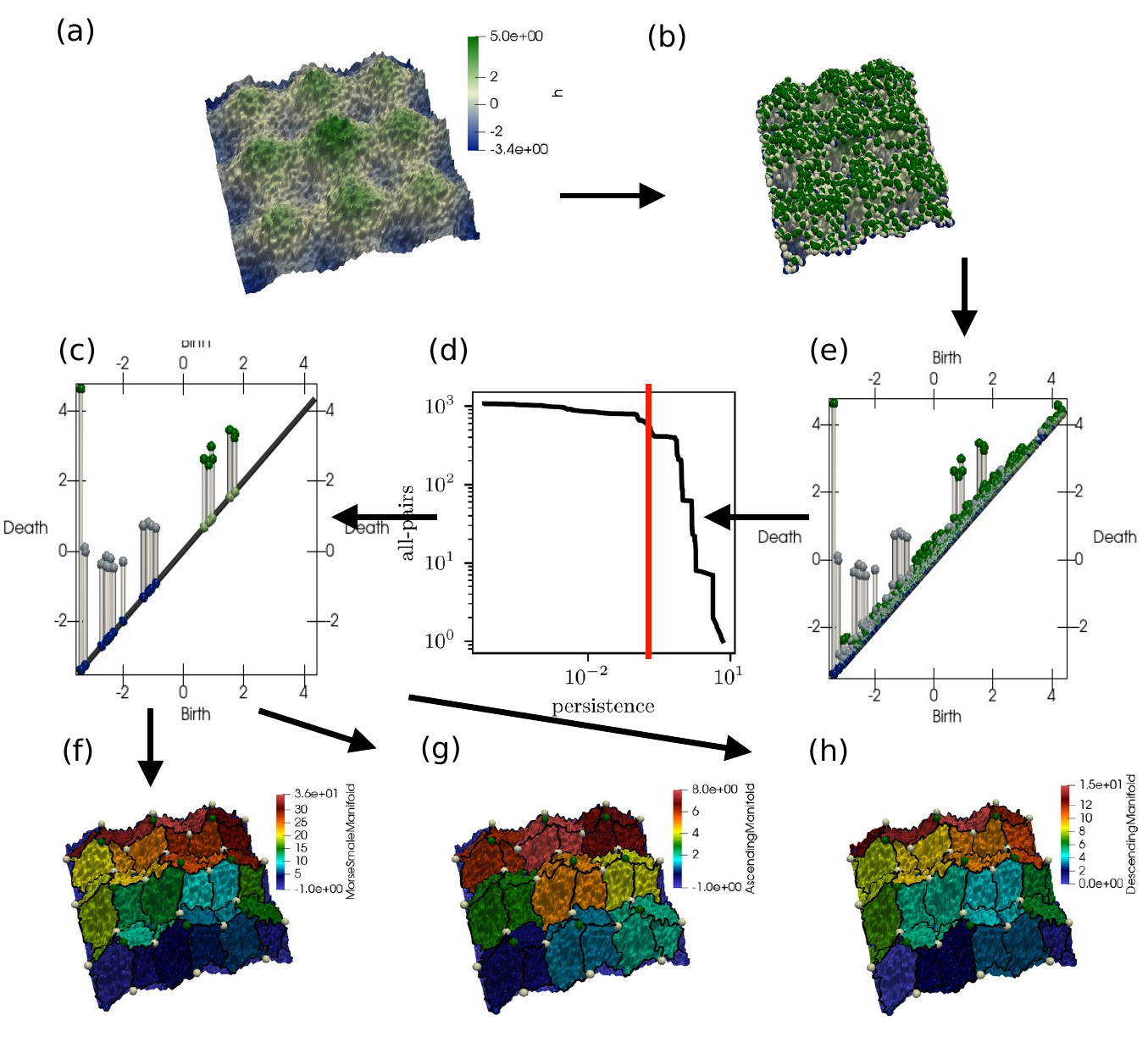}%
\caption{Morse-Smale segmentation example of a noisy terrain taken from \citep{ttk}: (a) shows the noisy surface $ f $ and (b) shows all the critical points  of the noisy surface from which we plot the persistence diagram $ \mathcal{P}(f) $ in (c) and persistence curve $ \mathcal{C}(f)$ in (d). (e) plots the persistence diagram after the application of a persistent threshold (red line in (d)). Once the noisy features are discarded, we get  the unique Morse-Smale segmentation (f) of the terrain. The same information can also be used to segment
the surface into unique ascending and descending manifolds as shown in (g) and (h) respectively. }%
 \label{fig:morse-manifold}%
\end{figure}

Let us consider an example from the documentation of the TTK library where the scalar field $ f $ is a discrete wavy terrain as shown in Fig. \ref{fig:morse-manifold} (a) where there are clear hills and valleys with some noise. If the critical points for this surface is calculated as shown by the spheres in Fig. \ref{fig:morse-manifold}(b), critical points are also located in all the noisy features of the terrain. The first step in TDA is to look at the distribution of critical points of $ f $ through the so-called \textit{persistence curve}, $\mathcal{C}(f)$, and the \textit{persistence diagram}, $ \mathcal{P}(f)$. For the wavy terrain shown, which is essentially a bivariate function, as we increase the function value, minima creates new isosurface components, maxima destroys components and saddle points merge or destroy components. Critical points can be arranged in a set of pairs, such that each critical point appears in only one pair $ (c_i, c_j)$ with $ f(c_i) < f(c_j)$ and $ \mathcal{I}(c_i) = \mathcal{I}(c_j) - 1$ \citep{edelsbrunner2022}. If one plots on the X-axis, $ f(c_i)$ and on the Y-axis a bar  with ordinates $ f(c_i)$ and $ f(c_j)$, then we obtain the \textit{persistence diagram}, $ \mathcal{P}(f)$, as shown in Fig. \ref{fig:morse-manifold} (e) and we call $ p = f(c_j) - f(c_i)$ as the \textit{persistence}. The noisy features of the terrain are \textit{less persistent} and hence created and destroyed (which can also be termed as birth and death) very quickly. On the persistence diagram, these features thus exist very close to the diagonal, while persistent features appear as much longer bars.  A persistence threshold can then be applied on this diagram (see Fig. \ref{fig:morse-manifold}(c)), to remove the noisy features, and evaluate the more significant features as shown by spheres in Fig. \ref{fig:morse-manifold} (f). Thus, the persistence diagram serves two purposes: a) it acts as a filtration technique to remove noisy features of the field and b) once the noisy features are discarded, the persistence diagram is a unique topological signature of the field, now barcoded into a two-dimensional plane. Finally, the number of critical pairs $ (c_i, c_j)$ can also be plotted
with the persistence $ $ resulting in the \textit{persistence curve}, $ \mathcal{C}(f)$, as shown in Fig. \ref{fig:morse-manifold} (d). Critical pairs at a low value of persistence that correspond to
the noisy features disappear quickly, and separate from the significant features through a horizontal plateau as shown by the red straight line in Fig. \ref{fig:morse-manifold}(d). 
Such a horizontal plateau is quite typical in analyzing scalar data as it clearly separates scales from smaller to larger features and can act as a good guide to decide the persistence threshold for applications. 

Once the noise is filtered, given a critical point $ p$, its \textit{ascending} ($\mathcal{A}(f)$ (resp. \textit{descending} $ \mathcal{D}(f)$) manifolds is defined as a set of points, belonging to integral lines (that are lines in the direction of $ \nabla f$) whose origin (resp. destination) is $ p$. In other words, ascending manifolds are segments where the tangent moves towards the local maxima, while descending manifolds move towards the local minima as shown in Figs. \ref{fig:morse-manifold}(g),(h). The transversal intersection of the ascending and descending manifolds
results in a unique Morse-Smale segmentation $ \mathcal{M}(f)$ as shown in Fig. \ref{fig:morse-manifold}(f). This methodology to uniquely segment a scalar function
into ascending and descending manifolds forms the main basis for our ideas on nanoporous materials. 

\subsection{Implementation}
Our \texttt{tda-segmentor} tool can be used to apply the aforementioned ideas on nanoporous structures such as, zeolites, MOF's or metallic nano-pillars. The code is open-source, developed in C++ 
and is published on \href{https://github.com/AMDatIMDEA/tda-segmentor}{GitHub}. The algorithms for topological analysis are taken from  the Topology Toolkit (TTK)
\citep{ttk}, which is a software platform that is easily accessible to the end users due to a tight integration with Paraview.
Moreover, the library comes with a variety of bindings in \texttt{Python} and \texttt{VTK/C++} for fast prototyping 
or even through direct \texttt{C++} without any \texttt{VTK} dependency. In our code, we take advantage of the
\texttt{VTK/C++} dependency as it results in much cleaner, efficient and better code.  

\begin{figure}[]%
\centering%
\includegraphics[width = 0.8\textwidth]{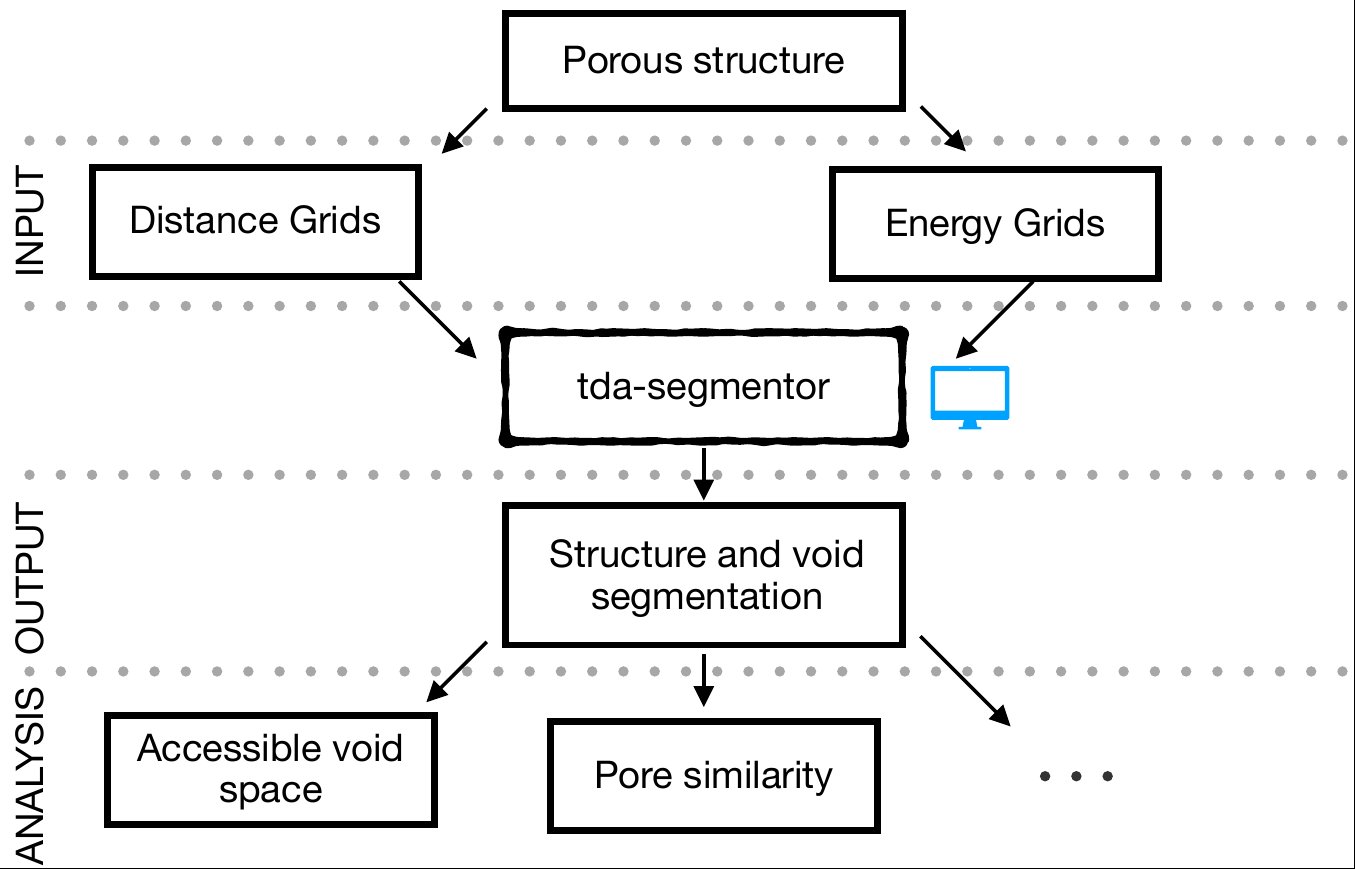}%
\caption{Workflow of the tool: From the porous structure, we first generate a 3D scalar function such as distance or energy grids. This is read into the segmentor tool which generates the segments which can then be used for analysis for applications. }%
 \label{fig:flow-chart}%
\end{figure}

Fig. \ref{fig:flow-chart} highlights the workflow of \texttt{tda-segmentor} tool. Starting from the porous structure, the first step is to generate a continuous scalar 
3-manifold that can be in the form of distance or energy grids. This acts as the scalar data input for the tool
to generate segments of void or solid structures.

\subsection{Inputs: Generation of scalar functions}
For our particular application to nanoporous structures, we generate two physically motivated volumetric scalar fields, i.e. the distance grids and the energy grids that are defined as follows. More generally, other similarly defined material datasets could be analyzed as well, e.g. electron densities, wave functions, guest molecule distributions etc.

\begin{figure}[ht!]%
\centering%
\includegraphics[width = \textwidth, trim = 0pt 0pt 0pt 30pt, clip = true]{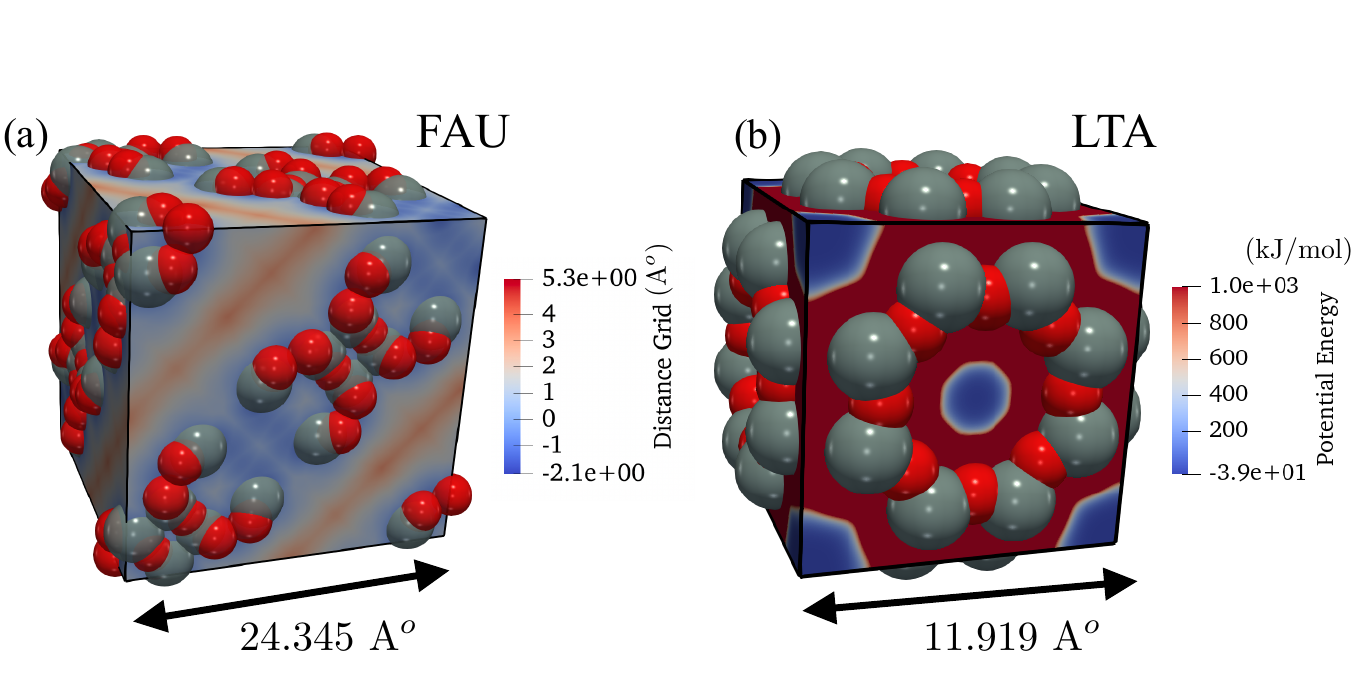}%
\caption{Two continuous scalar fields (a) distance grids for the FAU zeolite (b) and energy grid for a methane molecule in LTA zeolite framework serve as an input for segmentation. The red spheres are Oxygen (O) atoms, while the gray spheres are Silicon (Si). }%
 \label{fig:input-files}%
\end{figure}

\subsubsection{Distance grids}
As various chemical properties such as adsorption of a molecule directly depends on the pore morphology, the distance of a point to the material internal surface provides useful connections between structure and property. Additionally, we can use the sign of the distance function to extend the same algorithmic framework to study not only porosity but also the corresponding material structures. Specifically, we define the distance grid to be the least distance to the surface of the structure and negative within the structure
while positive in the pore (see Fig.~\ref{fig:input-files}(a) for the zeolite FAU and Fig. \ref{fig:gold-segmentation}(b) for gold nano-pillar structures). For zeolites, 
the distance grids can be evaluated using \texttt{Zeo++}\citep{Willems2012} that accepts common crystal structure input formats such as \texttt{.cif} and \texttt{.cssr}. 

\subsubsection{Energy grids}
When a guest molecule enters the porous materials, it faces repulsive and attractive forces resulting in a complex energy landscape that dictates, together with other parameters such as temperature,  
how the molecule diffuses and/or adsorbs within the pore network. The guest molecules'  binding sites - where by binding sites we mean locations where guest molecules are bound by physisorption - correspond to the minima of the energy landscape. Interaction energy grids (to be referred to as simply energy grids)  can be evaluated
from an existing open source repository \texttt{PorousMaterials.jl}\citep{PorousMaterials} that may use fairly standard Lennard-Jones potential based on 
the universal force field. An example for an energy grid is shown in Fig. \ref{fig:input-files} (b), for a methane
(CH4) molecule into a LTA zeolite framework.

Finally, periodic boundary conditions are applied to the volumetric grids by transforming any general lattice to an orthorhombic lattice. We note that for highly oblique triclinic lattices, such an approach might result in the minimum distance between two points measured incorrectly as the 
point that is closest could be in the second periodic image from the unit cell (see Chapter 3, Appendix C of \citep{Wassenaar2006}). We have included in the tool the option to consider the supercell by the command \texttt{-usesupercell} for such lattices, that constructs a $2 \times 2 $ unit cell on which periodic distances are then calculated correctly. 

%%%%%%%%%%%%%%%%%%%%%%%%%%%%%%%%%%%%%%%%%%%%%%%%%%%%%%%

\section{tda-segmentor: Usage and Examples}
\label{sec:tda-segmentor}

\begin{figure}[ht!]%
\centering%
\includegraphics[width = \textwidth]{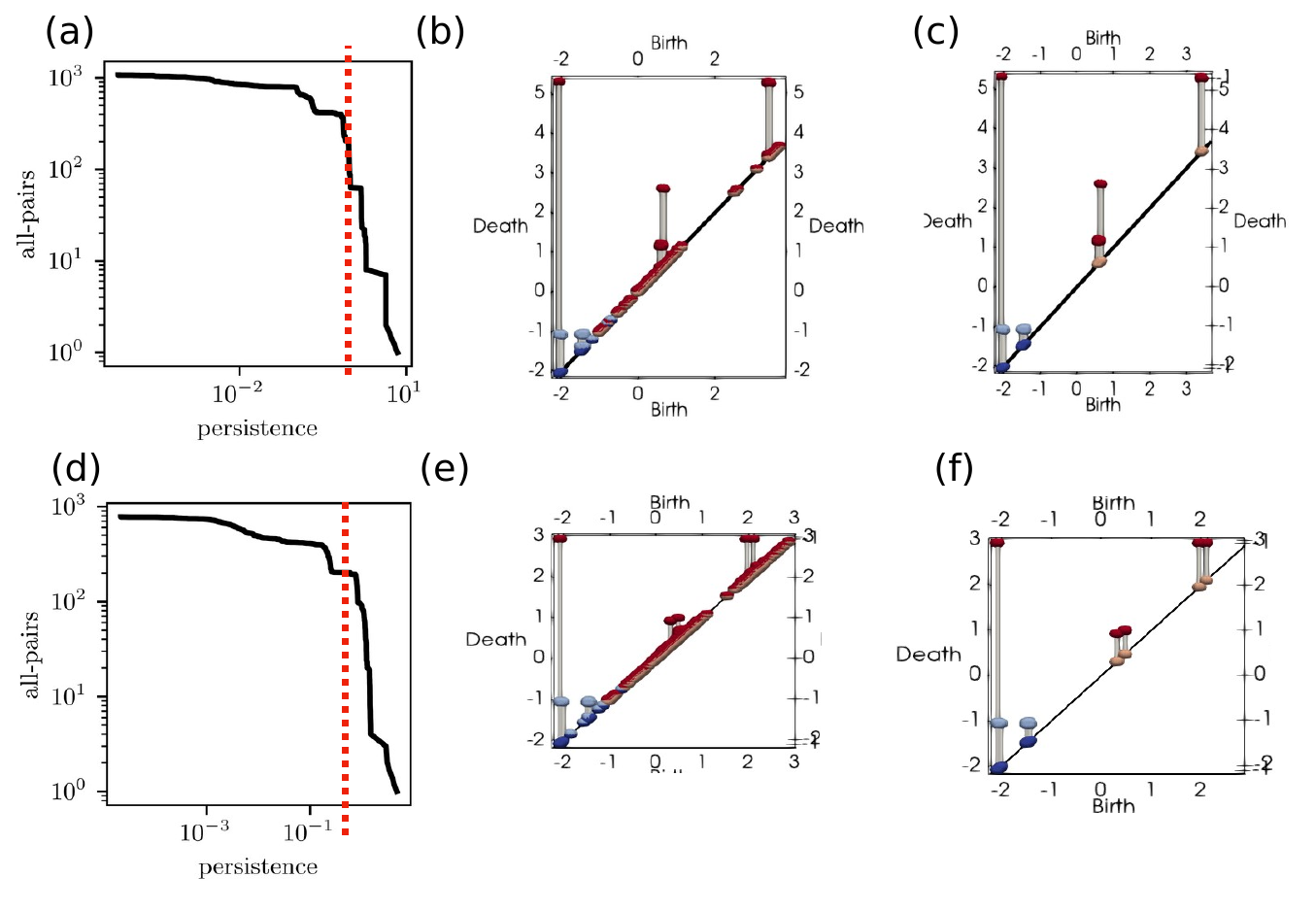}%
\caption{For the distance grid, $ f$, (a), (b), (c) plot the persistence curve, $ \mathcal{C} (f)$, persistence diagram $ \mathcal{P}(f)$ and persistence diagram $\mathcal{P}(f)$ after the application of a threshold for the zeolite FAU respectively, while the same is plotted for MFI in (d), (e) and (f) respectively. }%
 \label{fig:zeolite-segmentation-1}%
\end{figure}

\begin{figure}[ht!]%
\centering%
\includegraphics[width = \textwidth]{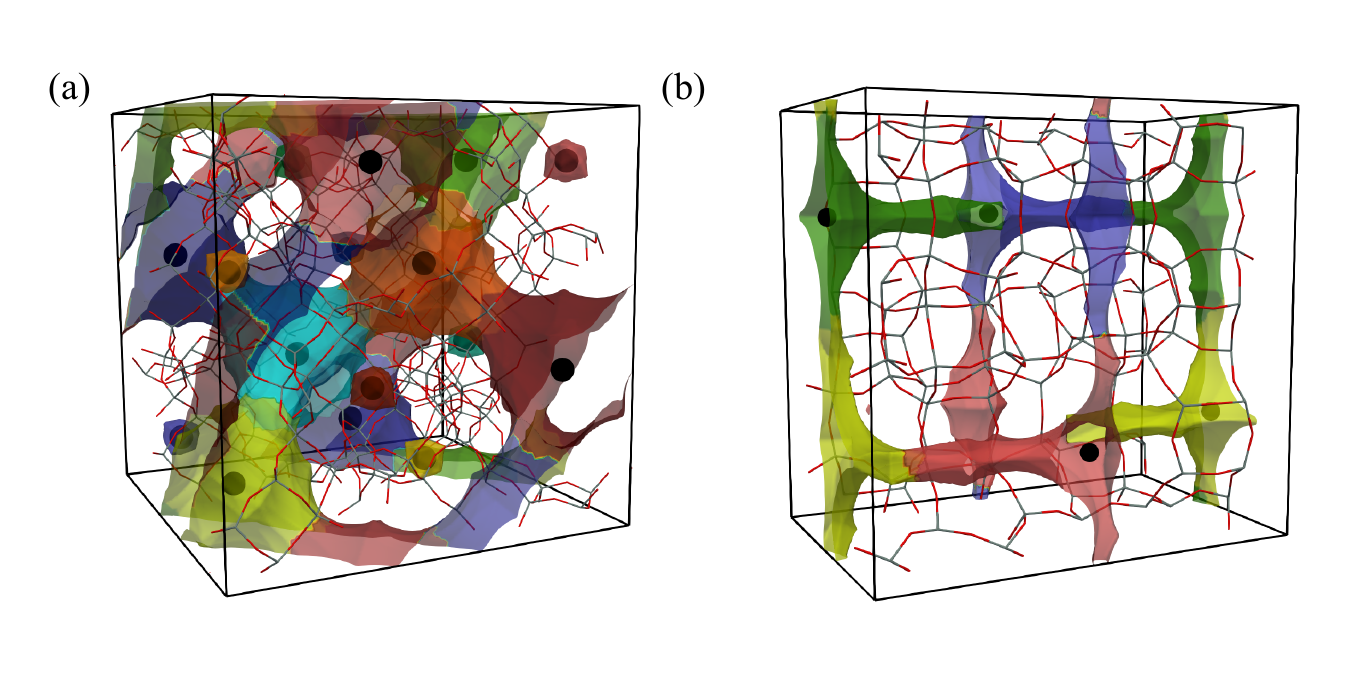}%
\caption{(a) and (b) show the void segmentation of the pore space that is accessible to a methane molecule for FAU and MFI respectively. 
Shown in black spheres in both (a) and (b) are the local maxima, that are located at the centre of the pores. }%
 \label{fig:zeolite-segmentation-2}%
\end{figure}

The \texttt{tda-segmentor} is written with a command-line invocation and different functionalities
of the code are implemented as modules, and detailed documentation is available on the GitHub repository. We explain these modules by applying them to a couple of well-known zeolites
such as FAU, MFI or LTA that have many potential applications as catalysts in the petroleum industry. We first generate 
the distance grid ($f$), which is shown in Fig. \ref{fig:input-files} (a) for the zeolite FAU and is done similarly for MFI. Once the distance grid is generated,
the persistence curve $ \mathcal{C}(f)$ is computed for both FAU and MFI as shown in Fig. \ref{fig:zeolite-segmentation-1}(a),(d) respectively. 
We clearly see that for both FAU and MFI, 
there is a clear plateau which separates smaller features from the larger features. Figs. \ref{fig:zeolite-segmentation-1} (b),(e)
then plots the persistence diagram ($\mathcal{P}(f)$ ) where noisy features are less persistent and populate close to the diagonal. By 
choosing a persistence threshold that corresponds to the plateau of the persistence curve $ \mathcal{C}(f)$, the noisy
critical points are removed resulting in persistence diagrams as shown in Figs. \ref{fig:zeolite-segmentation-1}(c),(f) respectively.

Once filtration is done by discarding critical points below a certain persistent threshold, Morse-Smale segmentation of the distance grid is performed
generating uniquely distinct segmented regions of the distance grid. As void regions are regions with positive distances that increase from the 
surface of the atoms, to analyze the void space we look at the ascending manifolds thresholded by a distance of 
$ r = 1.6 $ \AA, which corresponds to the void space accessible to a methane molecule. These segments for FAU and MFI
are shown in Fig. \ref{fig:zeolite-segmentation-2}(a) and (b) respectively, colored by their segment ID. Also shown as a black sphere at the center 
of the segment, is the local maxima of the distance grid. We would also like to add if the probe molecule is non-spherical, then the necessary segments for the major and minor radii can be extracted to be post-processed. 
 
 \begin{figure}[ht!]%
\centering%
\includegraphics[width = \textwidth, trim = 0pt 0pt 0pt 0pt, clip = true]{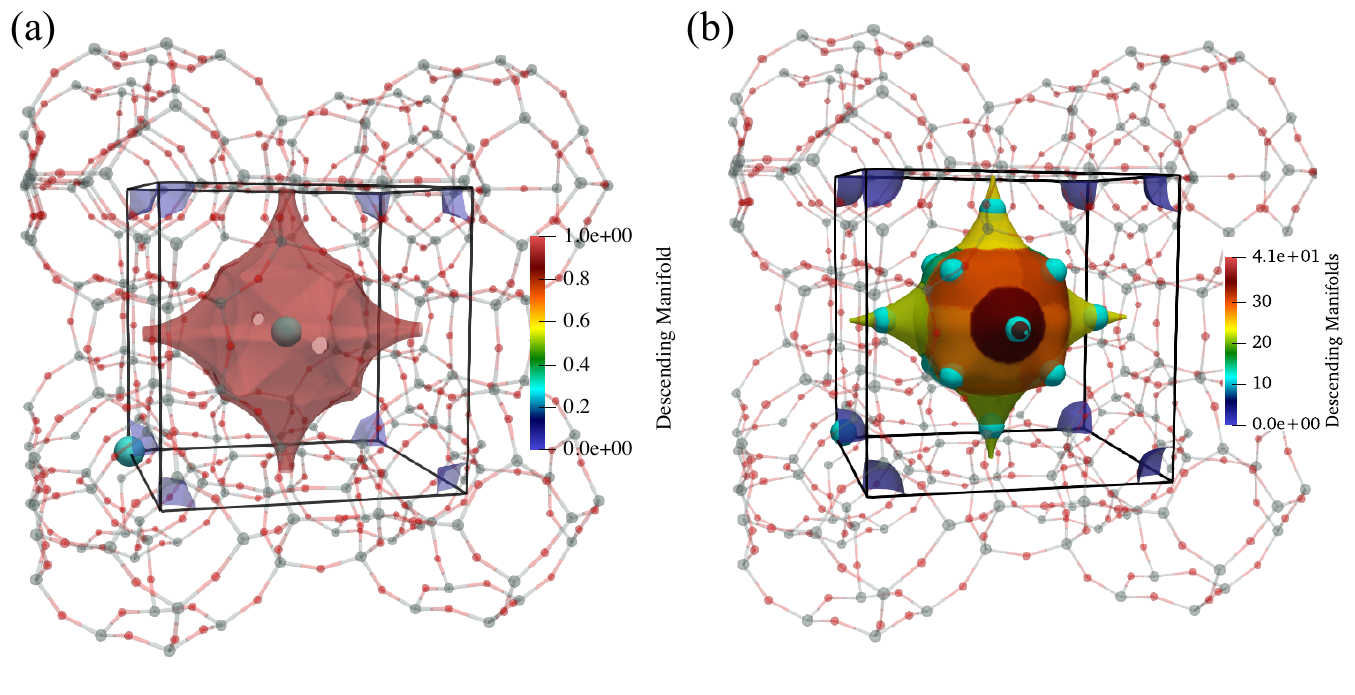}%
\caption{(a) shows the segmentation of void space accessible to a methane molecule, and clearly the distance based segmentation divides the void space into 
regions - a large accessible central pore and an inaccessible pocket in the corner. The two local maxima of these segments are shown in cyan spheres. (b) shows the segmentation obtained from the energy grid that segments the pore into many parts with the minima at the center which could be potential binding site for LTA.}%
 \label{fig:zeolite-segmentation-diff-inputs}%
\end{figure}

We now perform a similar analysis using energy grids and compare the results obtained from the segmentation of the distance grid. While the maxima 
and minima of the distance can correspond to the maximum pore and center of the atom respectively, local methane-probe interaction energy minima correspond to binding sites 
of the molecule in the pore. Fig. \ref{fig:zeolite-segmentation-diff-inputs} (a) presents the segmentation of the accessible pore for a methane molecule using the distance grid while Fig. \ref{fig:zeolite-segmentation-diff-inputs} (b)
visualizes the segmentation on an energy contour obtained from the energy grid. In (a), the void space is simply segmented into two regions, a large central pore,
and an isolated pore in the corner. 
The energy grids however give different information and segment the energy grid around the binding sites as shown by the cyan spheres in Fig. \ref{fig:zeolite-segmentation-diff-inputs}(b). 

\begin{figure}[ht!]%
\centering%
\includegraphics[width = \textwidth]{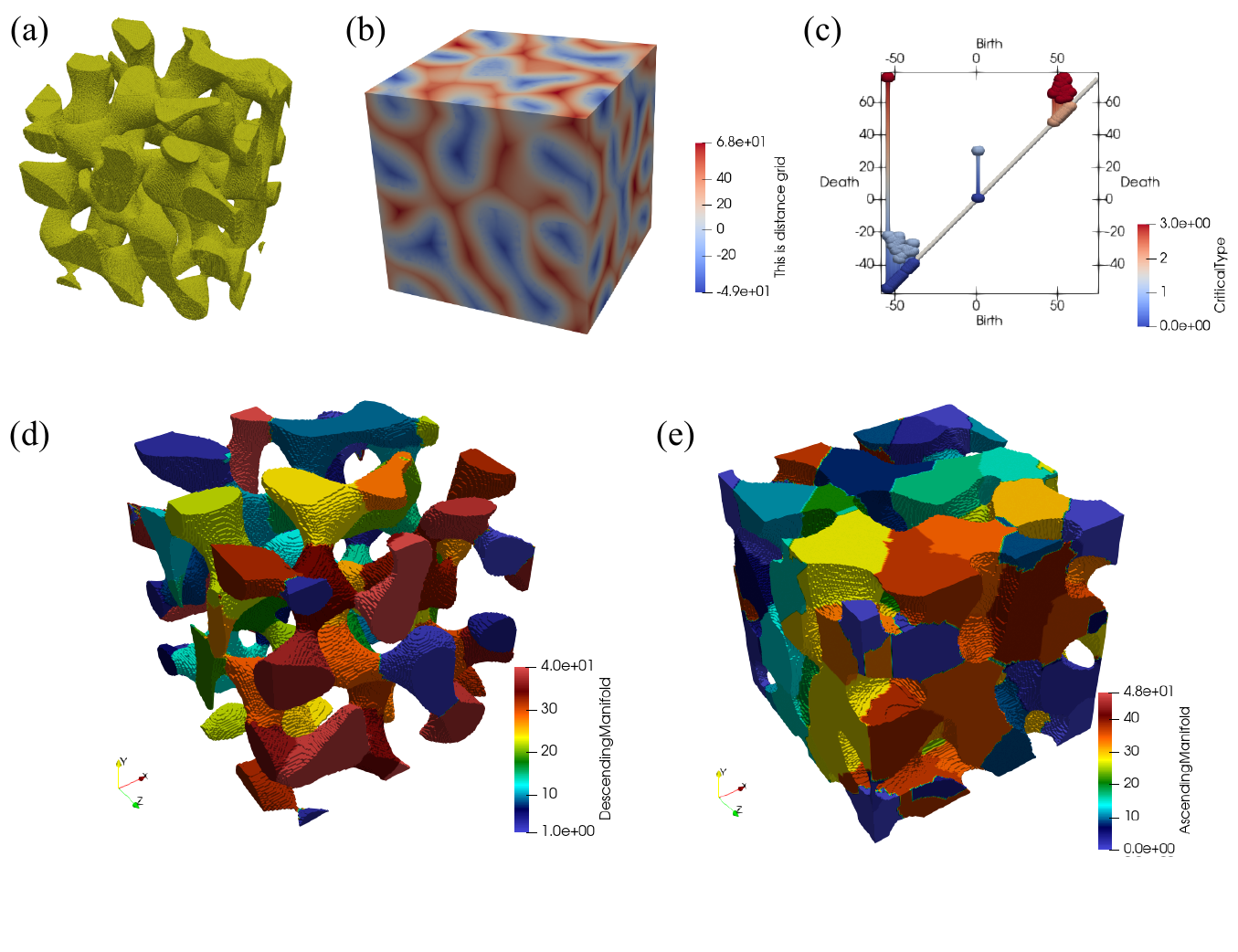}%
\caption{(a) is the gold nanoporous structure generated using random trigonometric fraction with a void fraction of 75 \%  (b) is its distance grid 
and (c) is the persistence diagram after the application of a certain threshold. By segmenting the ascending and descending manifolds respectively, we
segment the structure and pore as shown in (d) and (e) respectively. }%
 \label{fig:gold-segmentation}%
\end{figure}

Next, we present another example applied to a gold nano-pillar structure as shown in Fig. \ref{fig:gold-segmentation}(a). The topology of the pores
such as the ligament size and diameter contribute significantly in the macroscopic properties of the material such as the elastic modulus. From the nano-pillar structure, 
we can generate distance grids and by segmenting decreasing or increasing values of the distance grids, we obtain the segmentation of the structure or pore 
as shown in Figs. \ref{fig:gold-segmentation}(d), (e) respectively. 

\begin{figure}[ht!]%
	\centering%
	\includegraphics[width = \textwidth]{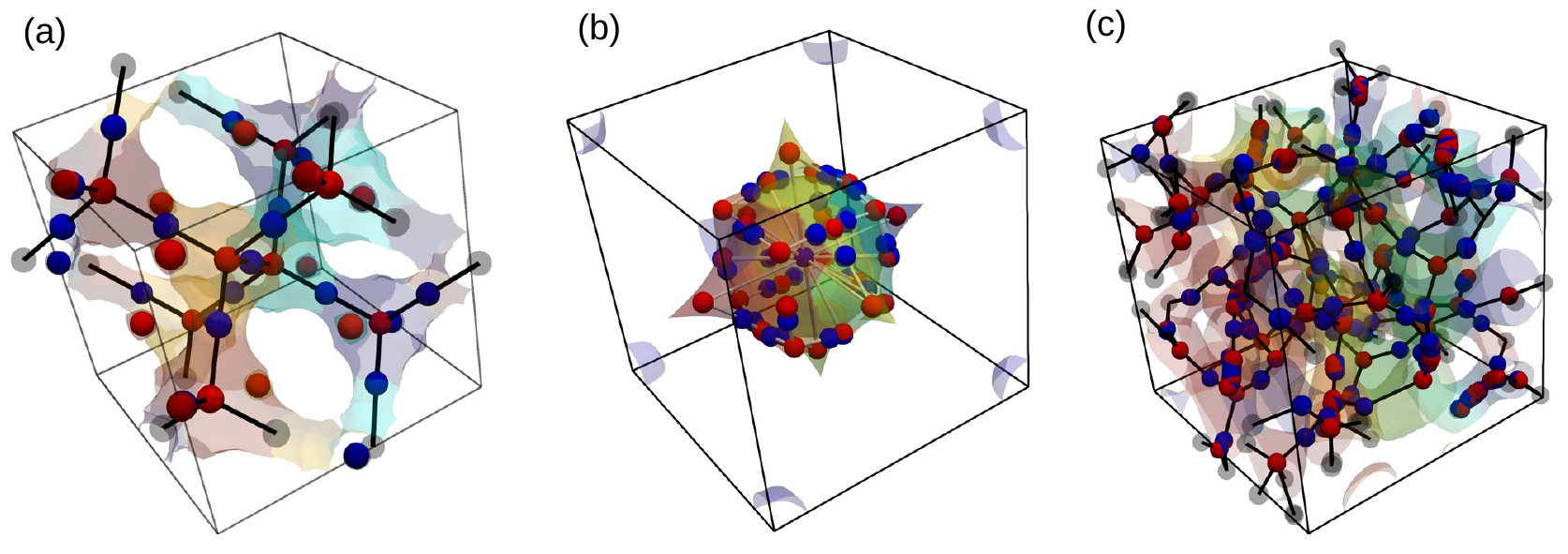}%
	\caption{Once the Morse-Smale complexes are evaluated, we can generate simple graph representations for nanoporous materials as shown in (a) for the void space accessible to a CH4 molecule in FAU zeolite where segmentation is done using distance grids. (b) shows the graph representation generated from the potential energy grids for a methane molecule in LTA zeolite while finally (c) shows the graph for the solid in gold nano-pillars. Show in red spheres are maxima in (a) while minima in (b) and (c) while red spheres are the corresponding 2-saddles (for (a)) and 1-saddles (for (b) and (c)) respectively. Gray spheres are nodes that are from the neighboring periodic box, shown for visualization. }%
	\label{fig:graph-representation}%
\end{figure}

Finally, once the Morse-Smale complexes are calculated, this information can be used to generate a graph structure for either the void space or the pore respectively. 
For the distance grids, maxima are located at the center of each segment and 
a 2-saddle is located at the border of neighboring segments, while for energy grids, minima are at the center of each segments and we will have 1-saddles at the border of each segment. By simply connecting the maxima/minima to their respective saddles, we can generate the graph representations for the void accessible to a guest molecule or the solid structure as shown in Fig. \ref{fig:graph-representation}.

%%%%%%%%%%%%%%%%%%%%%%%%%%%%%%%%%%%%%%%%%%%%%%%%%%%%%%%

\section{Application to nanoporous structures}
\label{sec:applications}

After the segmentation of the structure, we now present some applications 
on how these segments can be used to develop new ideas in the design and 
analysis of these structures. 

\subsection{Structure-property relations for zeolites}

\begin{figure}[ht!]%
\centering%
\includegraphics[width = \textwidth]{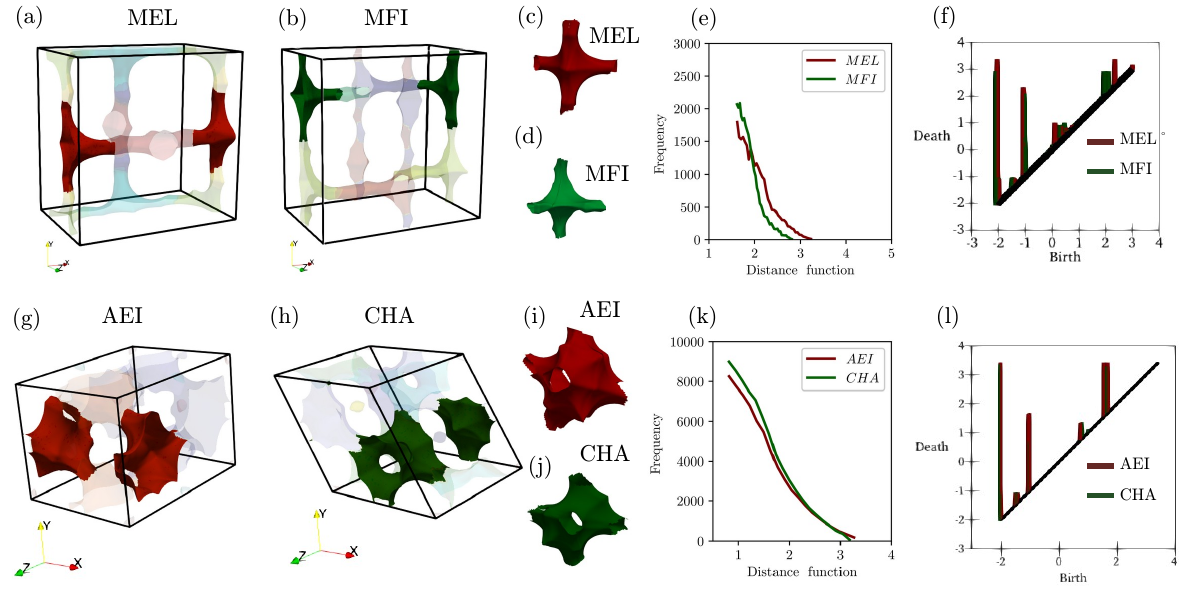}%
\caption{Topological methods to asses the similarity between zeolites: (a) and (b) show the accessible void space to a methane molecule of radius 1.6 \AA ~for two similar zeolites MEL and MFI respectively. These zeolites have similar pore segments (c), (d), similar distance histograms of only the segments (e) and very similar topological signature, given by their persistence diagrams (f). Similarity between another two zeolites AEI and CHA are assessed in (g)-(l) respectively. }
\label{fig:pore-similarity-zeo}
\end{figure}

Figure~\ref{fig:pore-similarity-zeo} shows two examples comparing MEL and MFI zeolite frameworks (Fig.~\ref{fig:pore-similarity-zeo} (a)-(f)) and AEI and CHA frameworks (Fig.~\ref{fig:pore-similarity-zeo} (g)-(l)) and how our tool can be useful in comparing these structures. MEL and MFI are both built from the same pentasil layers, which are the fundamental building blocks of these materials. However, in the MEL framework they are arranged through mirroring while they are arranged through inversion in MFI. This results with them having very similar properties, but with a straight channel system for MEL and a zig-zag channel system for MFI. Fig.~\ref{fig:pore-similarity-zeo}(a) and Fig.~\ref{fig:pore-similarity-zeo}(b) shows the segmentation accessible to a methane molecule for both MEL and MFI respectively and we see that the segment shapes are very similar for both the frameworks (Fig.~\ref{fig:pore-similarity-zeo}(c) and (d)) and hence the segment histograms (Fig.~\ref{fig:pore-similarity-zeo}(e)) are also similar. Likewise, for zeolites AEI and CHA,  that are built from double 6 rings (D6R) and are connected through mirror symmetry in AEI and slightly differently in CHA, and they also show similar pore segments. Moreover, by using only the segment information, we only use part of the structure for comparison while traditional approaches consider histograms of the entire structure which do not include the local pore topology which is absolutely crucial to understand the global adsorption property of the structure. Additionally, we can also compare the persistence diagrams of these zeolites, and zeolites with similar pore topology will show similar persistence diagrams as can be seen in Fig.~\ref{fig:pore-similarity-zeo} (f) and (l) respectively. Thus, our tool creates two sets of topological descriptors based on TDA that capture local pore geometry, viz: 1) the Morse-smale segments, where zeolites with similar segments will correlate with similar properties and 2) the persistence diagram which encodes all features, both local and larger features and similar structures can be identified by defining some normed distance (such as Hausdorff/Wasserstein distance) between their persistence diagrams.

\subsection{Inaccessible pore volume}

In section \ref{sec:tda-segmentor}, we showed some examples of the segmentation of the void space for a methane molecule of radius $ r = 1.6$ \AA (see Figs. \ref{fig:zeolite-segmentation-2} and \ref{fig:zeolite-segmentation-diff-inputs}(a)). If we look closer at the segments that are generated for FAU for example, some of the segments
are completely isolated and disconnected. These are regions that are 
inaccessible to a guest molecule which is important property to know in the
design of zeolites. We remind that once the Morse-Smale complexes are generated,
each Ascending Manifold will have a maxima within the segment, bordered by 2-saddles between two segments. By simply looking at the 2-saddles, we can identify these inaccessible pockets as such regions will not have any 2-saddles connected to any other segment. In our tool, if one calls the accessible void 
space module, segment information along with the connectivity of the segments
are saved which can be used to analyze these regions (see the documentation on the GitHub repository for more details). 
 
\subsection{Binding sites for a guest molecule}
From segmentation of energy grids as shown in Fig. \ref{fig:zeolite-segmentation-diff-inputs}(a), we can identify the locations that are local energy minima which can act as potential binding sites for a guest molecule such as methane. Moreover, we can also save information of the saddle points that lie on the border of two segments which can give us
the energy barrier that is required to translate a guest molecule from one binding site to another. 

\subsection{Structure-property relationship for Au nano-pillars}

\begin{figure}[ht!]%
\centering%
\includegraphics[width = \textwidth]{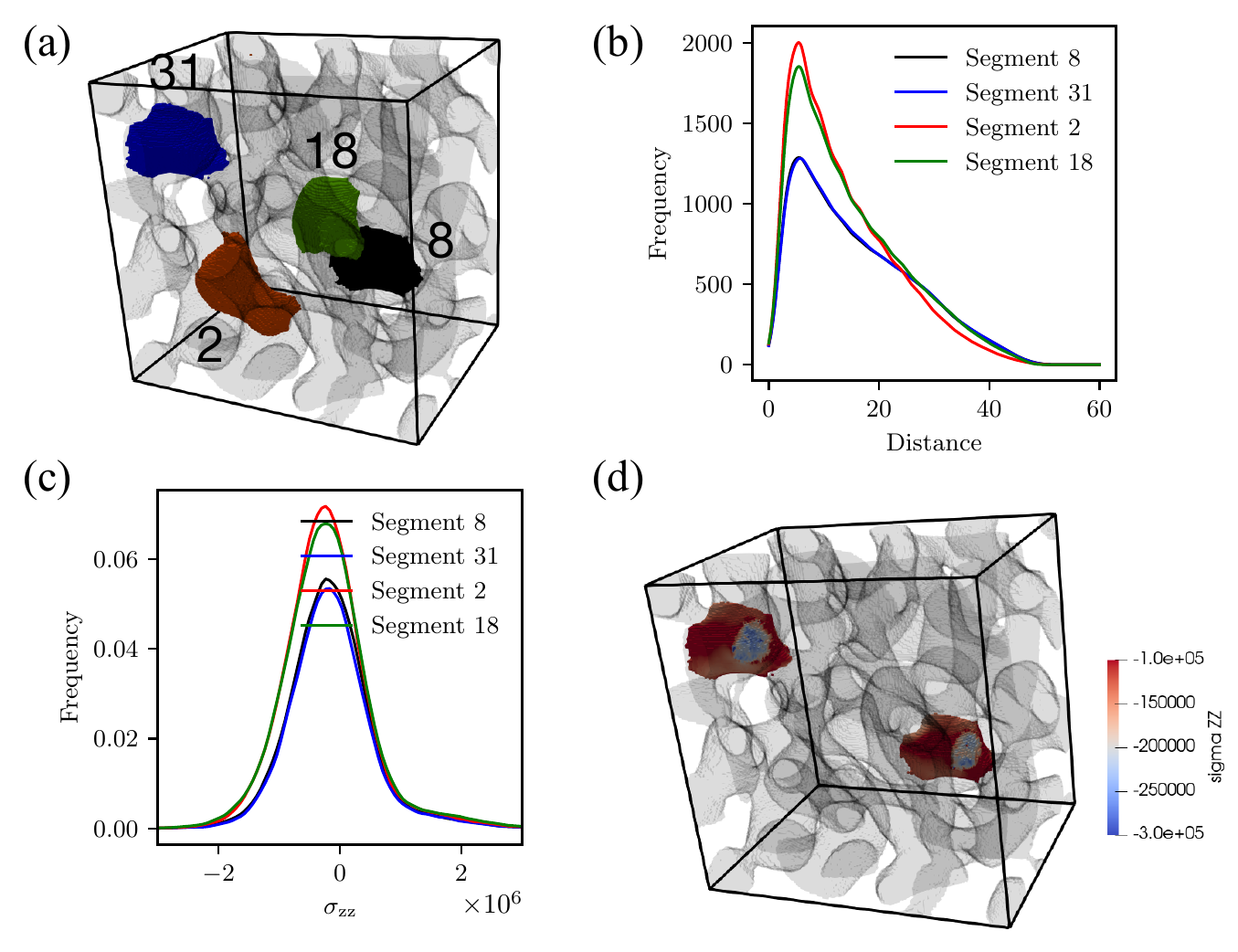}%
\caption{We analyze four segments 8, 31, 2 and 18 of a $15\%$ compressed gold nano-pillar as shown in (a). Segments that have similar distance histograms (see (b)) are
self-similar and also have similar stress histograms (c) and stress profiles (d). }
\label{fig:pore-similarity}
\end{figure}

We finally discuss a structural example where the gold nano-pillar shown in Fig. \ref{fig:gold-segmentation} is compressed by $15\%$. We generate the segments of 
the gold structure and compare the distance function histograms of each segment. Segments that have similar histograms are grouped together and are expected
to have similar properties. Fig. \ref{fig:pore-similarity} for example shows four segments grouped into two groups that have similar distance histograms (see Fig. \ref{fig:pore-similarity} (b)). These groups have similar stress histograms (see Fig. \ref{fig:pore-similarity} (c)) and similar stress profiles ( see Fig. \ref{fig:pore-similarity}(d)) thus correlating
structure with the geometry of the structure. 

%%%%%%%%%%%%%%%%%%%%%%%%%%%%%%%%%%%%%%%%%%%%%%%%%%%%%%%

\section{Conclusions}
\label{sec:conclusion}

In this article, we have presented a tool that offers a new line of analysis of porous structures taking inspiration from the growing field of topological data analysis. 
The presented tool takes as input either the distance or energy grids of the structure and can segment both the pore and the structure. Next, from the methods available in persistent homology such as plotting the persistence curves and persistence diagrams, we filter the noise from the scalar input data to segment only the most significant features. Finally we show a number of illustrative examples where the generated segments are analyzed to understand better correlations between the geometrical features and property. 
Furthermore, larger repositories of structures can to be analyzed with our code facilitating tasks such as database screening, structure diversity analysis and training of machine learning models.  

\section*{Acknowledgements}
\label{sec:acknowledgement}

We acknowledge the financial support from M-ERA.NET's PORMETALOMICS project supported by MCIN/AEI/10.13039/501100011033 and the European Union's NextGenerationEU/PRTR funds.

\section*{Data Availability}
The code is published as open-source and can be accessed here : \\ \href{https://github.com/AMDatIMDEA/tda-segmentor}{https://github.com/AMDatIMDEA/tda-segmentor}

\bibliographystyle{elsarticle-num} 
\bibliography{nanopore}

\end{document}